# Discovery of localized states of Fe 3d electrons in $Fe_{16}N_2$ and $Fe_8N$ films: an evidence of the existence of giant saturation magnetization

Xiaoqi Liu[1], Yun-Hao Xu[1], Cecilia Sanchez-Hanke[2] and Jian-Ping Wang[1*]

1. The Center for Micromagnetics and Information Technologies (MINT) and Department of Electrical and Computer Engineering, University of Minnesota, Minneapolis, MN, 55455

2. National Synchrotron Light Source, Brookhaven National Laboratory, Upton, NY 11973-5000

[*]Corresponding author: Email: jpwang@umn.edu; Tel: 612-625-9509

Abstract

The mystery of giant saturation magnetization of $Fe_{16}N_2$ has remained for 37 years. In this letter, X-ray absorption spectroscopy (XAS) and magnetic circular dichroism (XMCD) are used to study the electron states of Fe atoms in $Fe_{16}N_2$ and $Fe_8N$ films. Localized Fe 3d electron states, which are not expected from current theories (models), are observed. The XMCD spectra and elemental hysteresis loops reveal the polarization of N atoms and the hybridization of N and Fe atoms. This discovery points to the origin of giant saturation magnetization in $Fe_{16}N_2$.

Confirmation and understanding the origin of giant saturation magnetization is essential to revolutionalize the exchange-spring type permanent magnets for any energy efficient applications and breakthrough the bottleneck for the writability of current magnetic heads for extremely high areal density recording up to several Terabit/in$^2$ without switching to heat assisted magnetic recording and bit patterned media. Currently the highest saturation magnetization value ($4\pi M_s$) in available magnetic materials is about 2.45 T (FeCo alloy), which was predicted by the band magnetism theory. The $\alpha''$-$Fe_{16}N_2$ thin film was noted for its giant saturation magnetization (2.9 T) thirty-seven years ago [1, 2], followed by many experimental studies [3-11] with controversial results. Sugita et al [5, 6] reported an average value of 3.2 $\mu_B$/Fe in MBE-grown $\alpha''$-$Fe_{16}N_2$ single crystal thin films. Okamoto et al [7] and Sun [11] et al reported 2.8 $\mu_B$/Fe in high-quality FeN films with partial $\alpha''$-$Fe_{16}N_2$ phase by substrate-bombardment-free non-traditional deposition processes. Takahashi et al [8] and several other groups [9,10] reported less than 2.4 $\mu_B$/Fe in FeN films by traditional magnetron sputtering process.

Theoretical calculations have been done on the electron band structures of Fe and N in $\alpha''$-$Fe_{16}N_2$ structure [12-18]. Common to all the calculations is the use of three types of Fe sites in the $\alpha''$-$Fe_{16}N_2$ structure as shown in Fig.1. One of the general assumptions for all calculations is that Fe 3d electrons in $Fe_{16}N_2$ structure are itinerary (delocalized) as in other magnetic metals or alloys. However, the calculated average magnetic moment per Fe atom has never been higher than 2.5 $\mu_B$/Fe. In this letter, we will first report the successful fabrication of FeN films with partial $\alpha''$-$Fe_{16}N_2$ phase, for which giant saturation magnetization (2.9 T) is confirmed. Then we will report the discovery of the localized states of Fe 3d electrons in FeN films with $Fe_{16}N_2$ and $Fe_8N$ structures, which hasn't been expected and applied for all previous theoretical calculations on $Fe_{16}N_2$ structure, by using X-ray absorption spectroscopy (XAS) and x-ray magnetic circular dichroism (XMCD) technique

[19, 20].

The FeN thin films with single $Fe_8N$ phase and $Fe_{16}N_2 + Fe_8N$ mixed phases were grown by sputtering Fe targets in a mixture gas of Ar and $N_2$ using a facing-target-sputtering system. The base pressure of the sputtering chamber was $2\times10^{-7}$ Torr. Si (100) substrates capping with about 100 nm $SiO_2$ were used. By adjusting the substrate temperature and the ratio of Ar and $N_2$ in the gas flow, a series of FeN films with different phases were successfully fabricated. The crystal structure of FeN thin films was characterized by a Co source x-ray diffractometer (XRD). Their magnetic properties were measured by a vibrating sample magnetometer (VSM).

Fig. 2 shows the XRD spectra of two 50 nm thick FeN films. For sample A, besides $SiO_2(213)$ and Si(400) peaks from the substrate, only one peak at 52.4° appears, corresponding to $Fe_8N(110)$ peak. To confirm the single $Fe_8N$ phase in sample A, $Fe_8N$ (101) and (002) peaks were observed in grazing incident angle XRD measurements. Together, these peaks exclude other FeN phases such as Fe [30], $Fe_2N$ [31], $Fe_3N$ [32], or $Fe_4N$ [32]. For sample B, three peaks are observed at 49.9°, 52.4° and 58.6°, corresponding to $Fe_8N(101)$, $Fe_8N(110)$ and $Fe_{16}N_2(301)$ respectively. The $Fe_{16}N_2(301)$ peak is one of characteristic peaks for $Fe_{16}N_2$ phase. Another characteristic peak, $Fe_{16}N_2(002)$, was detected by grazing incident angle XRD measurement. Therefore sample B contains two FeN phases, $Fe_8N$ and $Fe_{16}N_2$. The saturation magnetization ($4\pi M_s$) of sample A and B measured by the VSM are 2.14 T and 2.39 T respectively. The $M_s$ value of $Fe_8N$ (sample A) agrees very well with that reported by Okamoto et al [33]. The volume fraction of $Fe_{16}N_2$ phase in sample B is estimated to be about 30% by considering peak intensities of $Fe_{16}N_2(301)$ and $Fe_8N(110)$ in the XRD pattern and the structure factors of these two peaks in the bct crystalline structure [34]. The saturation magnetization ($4\pi M_s$) of $Fe_{16}N_2$ phase was therefore calculated to be 2.96 T, which agrees well with that reported by Sugita et al [6].

The XAS and XMCD were performed on beamline X13A of the National Synchrotron Light Source (NSLS) at Brookhaven National Laboratory (BNL). The XAS and XMCD spectra were recorded in total electron yield mode (TEY) using elliptically polarized soft x-rays with ~70% degree of circular polarization. A magnetic field up to 0.2 T was applied parallel to the sample surface and along the incident x-ray direction. The photon incident angle was set at 8 degree. The line structures of XAS and XMCD directly represent the Fe 3d state electronic configurations [21, 22]. The spin and orbital moment of Fe atoms are studied based on XMCD sum rules [23, 24, 25]. The XAS and XMCD spectra with different multiplet structures are used to identify different localized Fe 3d states, as demonstrated in theory and experiment [26 - 29].

The remarkable observation in the XAS and XMCD spectra of $Fe_8N$ and $Fe_{16}N_2$ + $Fe_8N$ films is the appearance of multiplet structures in the energy range of the Fe $L_{3,2}$ edge, as shown in Fig 3. The mutiplet positions in the spectra were labeled with black dashed lines in fig 3 (A) and (B). Published XMCD spectra from bulk iron [25] and localized Fe $3d^7$ state [29] were also included in Fig.3 for comparison. In XAS and XMCD spectra, Fe $L_{3,2}$ edges with sharp multiplet structures are normally fingerprints of localized Fe atomic configurations [28, 29]. The reference XMCD spectra of delocalized Fe 3d states in bulk iron (Fig. 3(C)) and localized Fe $3d^7$ state (Fig. 3(D)) indicate that only XMCD spectra of localized Fe 3d states contain multiplet Fe $L_{3,2}$ edges. In the XMCD spectra of delocalized Fe 3d states, multiplet Fe $L_{3,2}$ edges are clearly absent, only broad single negative $L_3$ and positive $L_2$ edges were obtained because of the presence of band like 3d states structure rather than discrete states. Therefore, these multiplet Fe $L_{3,2}$ edges in the XAS and XMCD spectra provide the first direct evidence of the existence of localized Fe 3d states in $Fe_8N$ and $Fe_{16}N_2$ structures.

Although localized Fe 3d states were observed in $Fe_8N$ and $Fe_{16}N_2$ films, the fact that $Fe_8N$ and $Fe_{16}N_2$ films possess low electrical resistivity, which is comparable to iron film [35], reveals that plenty of itinerary electrons still exist in the $Fe_8N$ and $Fe_{16}N_2$ structures. This implies that these localized Fe 3d states are not completely but partially localized.

Furthermore, Fe 3d states in $Fe_8N$ and $Fe_{16}N_2$ structures are confirmed to be partially localized based on a quantitative indicator, ratio R, defined as the ratio between Fe $L_2$ peak area and $L_3$ peak area in the XMCD spectra. Based on the equations of sum rules, the orbital moment contribution to the total magnetic moment of Fe atoms can be deduced from the R value [36]. An R ~ –1 indicates completely quenched orbital moment ($M_{orbital}$ ~ 0), which is a basic characteristic of delocalized Fe state. For localized Fe 3d states, the orbital moment can be completely or partially unquenched. So the R value will be deviated from –1. As summarized in table 1, R values for different Fe 3d states were extracted from XMCD spectra based on the method presented by Brewer et al [36]. The R values for sample A and sample B were – 0.76 and – 0.72 respectively, which indicate partially unquenched Fe 3d orbital moments, since the R values for localized Fe $3d^5$, $3d^6$ and delocalized Fe 3d states are – 1, – 0.36 and – 0.90 respectively. This result strongly supports the argument that Fe states in sample A and B are partially localized Fe 3d states. Therefore, the R values for sample A and B are strong evidence of the existence of localized Fe 3d states in $Fe_8N$ and $Fe_{16}N_2$ structures. The differences between the XMCD spectra and R values of sample A and B also suggest that Fe 3d states in $Fe_{16}N_2$ are different from that in $Fe_8N$.

Fe atoms in bcc bulk iron are averaging 6.61 3d electrons. Therefore in bct $Fe_8N/Fe_{16}N_2$ structure, the 3d electron numbers of Fe atoms will be in the range of 5-7 electrons. Table 1 shows the magnetic moments for localized Fe $3d^5$, $3d^6$ and $3d^7$ states which are calculated based on Hund' rules. All three localized Fe states possess much enhanced total magnetic moment compared to delocalized Fe states (in bulk iron). The localization of Fe 3d states largely enhances both the orbital moment (except $3d^5$ state) and spin moment of Fe atoms. The same moment enhancement effect was also reported by Gambardella et al on isolated Fe atoms with $3d^7$ configuration [29]. The finding of localized Fe 3d states with enhanced magnetic moments in $Fe_{16}N_2$ structures points to the origin of giant saturation magnetization of $Fe_{16}N_2$, considering that the unit cell volume of bct-$Fe_{16}N_2$ is just 10% larger than bcc-Fe structure.

One immediate question is what may cause the loss of giant saturation magnetization in $Fe_8N$ structure in which localized Fe 3d states were also observed. One possible reason could be the lack of ordering of N atoms in $Fe_8N$ structure. The random distributed N atoms in $Fe_8N$ will result in a decrease in the number of Fe atoms in localized states and an increase in the number of Fe atoms in delocalized states, which both can lead to the decrease of total magnetization of $Fe_8N$.

In bct $Fe_{16}N_2$ and $Fe_8N$ structures (figure 1 (a) and (b)), the localization of Fe 3d states from the delocalized Fe 3d state in bcc bulk iron arise from the perturbation of the crystal field of N atoms. Since Fe(4e) and Fe(8h) site are 60% and 50% closer to the N sites in distance than Fe(4d) site (Fig.1 (c-e)), Fe atoms at Fe(4e) and Fe(8h) sites should have stronger bonding with neighbor N atoms than that at Fe(4d) site, which indicate that Fe atoms at Fe(4e) and Fe(8h) sites have higher degree of localization than at Fe(4d) site.

The bonding (hybridization) between Fe atoms and N atoms is proved by a further XMCD analysis that was performed in the energy range of 340-460 eV for sample A and B. The Nitrogen K edge is found at 397.0 eV, which is consistent with the XPS result (~398eV) [37]. The N K edge moves to lower energy compared to the binding energy of 1s electrons of single N atom (409.9 eV) [38]. The hybridization of N atoms with partially localized Fe atoms is expected to cause the N atoms negatively charged. Therefore the Coulomb interaction between electrons results in the increase of energy level of N 1s states and N K edge shifting to lower energy.

Hysteresis loops of N and Fe atoms recorded at the N K edge and Fe $L_3$ edge with an in-plane magnetic field provide direct evidence of the spin polarization of N atoms and the hybridization of N and Fe atoms as shown in Fig. 4. The hysteresis loops of N and Fe atoms exhibit the same coercive field of 15 Oe for both sample A and B. The spin polarization of N atoms and the hybridization between N and Fe atom could be the main cause of the localization of the 3d states of Fe atoms in the $Fe_8N$ and $Fe_{16}N_2$ structures [39].

In conclusion, we report the discovery of partial localized Fe 3d states in $Fe_{16}N_2$ and $Fe_8N$ films. The partial localization of Fe 3d states introduced large unquenched orbital moment and enhanced spin moment in Fe atoms, therefore greatly increased the average magnetic moment of Fe atoms. This discovery may not only explain the origins of giant saturation magnetization in $Fe_{16}N_2$ structure but also points out a direction to search for new magnetic materials with giant saturation magnetization.

The work was partially supported by the U.S. Department of Energy, Office of Basic Energy Sciences under contract No. DE-AC02-98CH10886, National Science Foundation NNIN program at University of Minnesota, Seagate Technology and Western Digital Cooperation. The authors would like to thank Prof. Frank De Groot of Utrecht University for his help on the XMCD multiplet calculations, the useful conversations with Mr. Nian Ji, Prof. Jack Judy, Dr. Mark Kief and Dr. Yinjian Chen.

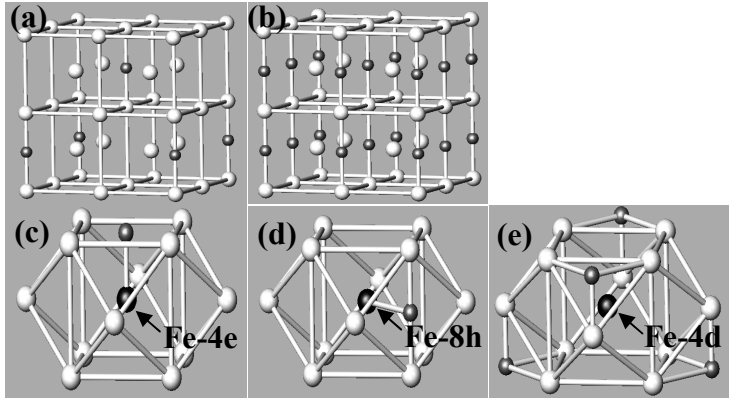

Fig. 1 Crystalline structure of (a) $Fe_{16}N_2$ and (b) $Fe_8N$ (white-Fe; grey-N), (For $Fe_8N$, N atoms randomly distribute at all the possible N sites and keep N-Fe atom ratio at 1:8); Crystalline environment of (c) Fe-4e; (d) Fe-8h; (e) Fe-4d.

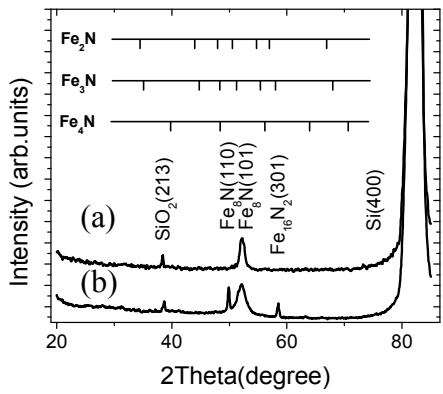

Fig. 2 XRD spectra of (a) sample A with single $Fe_8N$ phase and (b) sample B with $Fe_8N+Fe_{16}N_2$ phases.

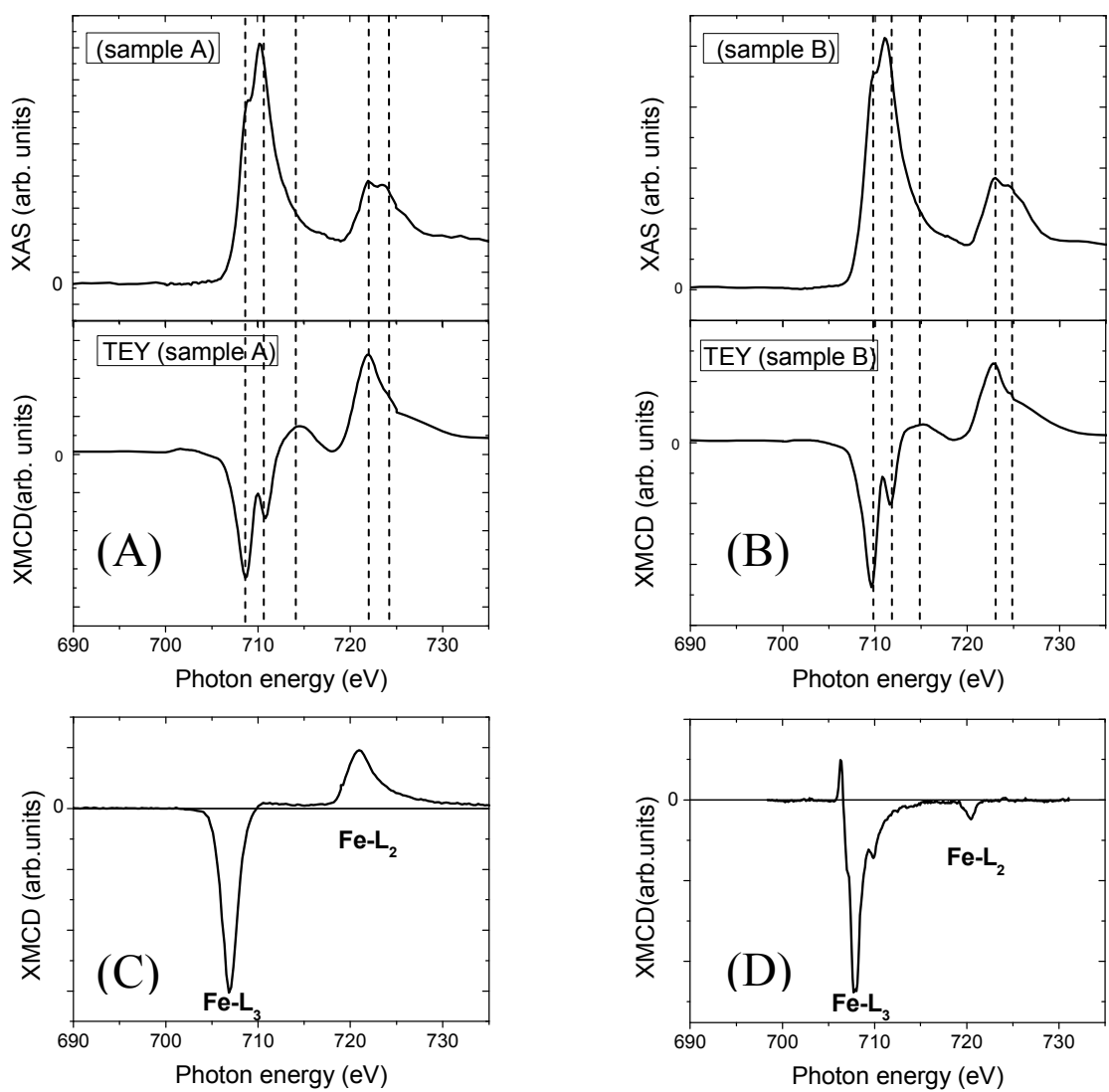

Fig. 3 XAS and XMCD spectra of (A) sample A, (B) sample B, (C) delocalized Fe 3d state in bulk iron [25] and (D)localized Fe $3d^7$ state [29].

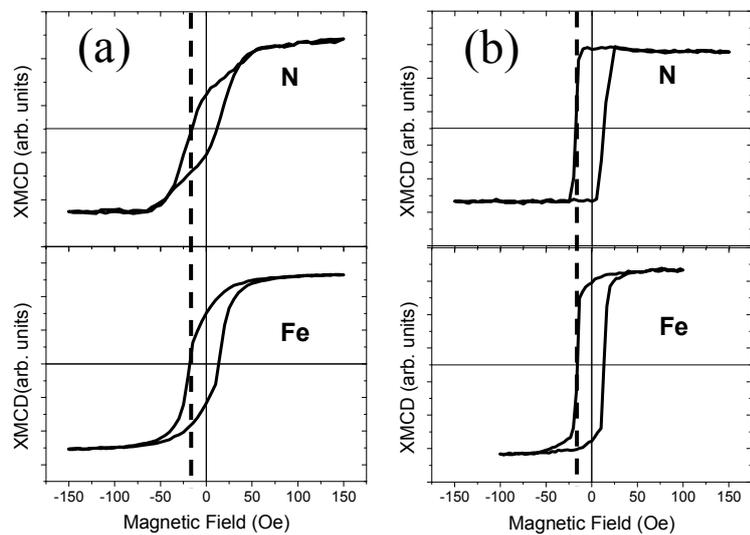

Fig. 4 Hysteresis loops recorded for N atoms at K-edge 399 eV and Fe atoms at $L_3$ edge 707eV on (a) sample A and (b) sample B.

Table 1 Ratio R, orbital ($M_o$), spin ($M_s$) and total magnetic moment (M) of different Fe 3d states, extracted from XMCD spectra in Fig. 3. (M value of $Fe_{16}N_2$ phase was calculated from its saturation magnetization value. R values of $Fe3d^5$ and $3d^6$ were extracted from the calculated XMCD spectra by G. van der Laan [28]. The magnetic moment of Fe $3d^5$, $3d^6$ and $3d^7$ were calculated by Hund's rules.)

|  | $Fe_8N$ (sample A) (experiment) | $Fe_{16}N_2+Fe_8N$ (sample B) (experiment) | $Fe_{16}N_2$ | Fe $3d^5$ [28] | Fe $3d^6$ [28] | Fe $3d^7$ [29] | Fe in bulk iron[25] |
|---|---|---|---|---|---|---|---|
| Ratio R | -0.76 | -0.72 |  | -1 | -0.36 | 0.09 | -0.90 |
| $M_o(\mu_B)$ |  |  |  | 0 | 2.45 | 3.46 | 0.09 |
| $M_s(\mu_B)$ |  |  |  | 5.92 | 4.90 | 3.87 | 1.98 |
| $M(\mu_B)$ | 2.20 | 2.46 | 3.20 | 5.92 | 6.71 | 6.67 | 2.07 |